\documentclass{Interspeech2024}
\usepackage{multirow}
\usepackage{caption}
\usepackage{threeparttable}
\usepackage{boldline} 

\interspeechcameraready

\title{Phonetic Enhanced Language Modeling for Text-to-Speech Synthesis}

\name[]{Kun}{Zhou}
\name[]{Shengkui}{Zhao}
\name[]{Yukun}{Ma}
\name[]{Chong}{Zhang}
\name[]{Hao}{Wang} 
\name[]{Dianwen}{Ng}
\name[]{Chongjia}{Ni}
\name[]{Nguyen Trung}{Hieu}
\name[]{Jia Qi}{Yip}
\name[]{Bin}{Ma}

\address{Alibaba Group, Singapore}
\email{kun.z@alibaba-inc.com}


\begin{document}

\maketitle

\begin{abstract}
Recent language model-based text-to-speech (TTS) frameworks demonstrate scalability and in-context learning capabilities. However, they suffer from robustness issues due to the accumulation of errors in speech unit predictions during autoregressive language modeling. In this paper, we propose a phonetic enhanced language modeling method to improve the performance of TTS models. We leverage self-supervised representations that are phonetically rich as the training target for the autoregressive language model. Subsequently, a non-autoregressive model is employed to predict discrete acoustic codecs that contain fine-grained acoustic details. The TTS model focuses solely on linguistic modeling during autoregressive training, thereby reducing the error propagation that occurs in non-autoregressive training. Both objective and subjective evaluations validate the effectiveness of our proposed method.

\end{abstract}

\section{Introduction}
Text-to-Speech (TTS) aims to produce natural and intelligible speech from text input \cite{dutoit1997introduction}. With the advent of deep learning methods, neural network-based TTS systems have significantly enhanced the quality of synthesized speech \cite{tan2021survey}. TTS has become a pivotal technology in conversational artificial intelligence \cite{gao2018neural} and human-machine interaction \cite{zhou2022speech}.

The primary challenge for text-to-speech systems is the computational modeling of the one-to-many mapping function between text and speech \cite{dutoit1997introduction}. A TTS model is expected to synthesize highly natural speech across various styles for multiple speakers, including unseen speakers or styles (zero-shot TTS) \cite{tan2021survey}. Previous TTS studies \cite{jia2018transfer,casanova2022yourtts,cooper2020zero,wu2022adaspeech} have largely focused on designing a speaker or reference encoder to capture speaker characteristics and incorporate them into a TTS framework. Since speaker characteristics are closely entangled with speech content and style \cite{zhou2021vaw,zhou2022emotion}, speaker similarity and speech naturalness often decline in zero-shot scenarios \cite{casanova2022yourtts}. Moreover, these frameworks require high-quality recorded speech data for training, which constrains the diversity of the synthesized speech.

Recent  language models have achieved great success in natural language generation \cite{achiam2023gpt, chang2023survey} and then applied to audio generation \cite{kreuk2022audiogen} and speech synthesis \cite{borsos2023audiolm, kharitonov2023speak}. Language model (LM) based TTS systems \cite{wang2023neural, zhang2023speak} formulate text-to-speech as a conditional language modeling task, where the speech waveform is quantized into discrete tokens and is then modeled with a language model. Compared to conventional neural TTS, language model-based TTS systems have demonstrated strong in-context learning capabilities on large and diverse speech data, and can synthesize highly natural speech. Despite their advancements, these systems often face issues with robustness and suffer from word deletion, repetition, and mispronunciation \cite{wang2023neural}, which makes them less ideal for real-life applications.

One notable challenge in text-to-speech systems based on language models lies in their autoregressive approach. For instance, in \cite{wang2023neural}, text tokens are mapped to speech tokens in an autoregressive manner. However, accurately modeling speech tokens is difficult due to the presence of both verbal and non-verbal information in the speech signal \cite{zhou2021emotional,zhou2020converting}. Furthermore, additional challenges arise from confounding factors such as environmental noises, which are encoded into speech tokens during quantization \cite{defossez2022high}. False predictions of speech tokens can accumulate, leading to robustness issues in synthesized speech. In this paper,
we introduce a phonetic enhanced language modeling method for TTS that leverages self-supervised learning (SSL) representations \cite{mohamed2022self} as the training target for autoregressive training.  This approach allows the TTS framework to focus solely on linguistic modeling during autoregressive training.  Subsequently, a non-autoregressive model is employed to recover fine-grained acoustic details of speech from the predicted SSL representations.
Our contributions are summarized as follows:
\setlength{\leftmargini}{10pt} 
{\begin{itemize}
\setlength{\itemindent}{0pt}   
    \item We propose a phonetic enhanced language modeling method to improve the performance of language model-based TTS;
    \item We leverage SSL representations, which are phonetically rich, as the training target for autoregressive language model training. We map text tokens to phonetic variations autoregressively, and phonetic variations to acoustic details in a non-autoregressive manner, thereby reducing the error propagation;
    \item Our method enhances the performance of TTS frameworks by effectively improving system robustness and achieving better speech quality in terms of naturalness and zero-shot speaker similarity compared to the baseline.
\end{itemize}}

The rest of the paper is organized as follows:  We introduce the related work on language model-based TTS in Section 2.
Our proposed method is introduced in Section 3. In Section 4, we report our experimental results. Section 5 concludes the paper.

\begin{figure*}[t]
    \centering
    \includegraphics[width=1\textwidth]{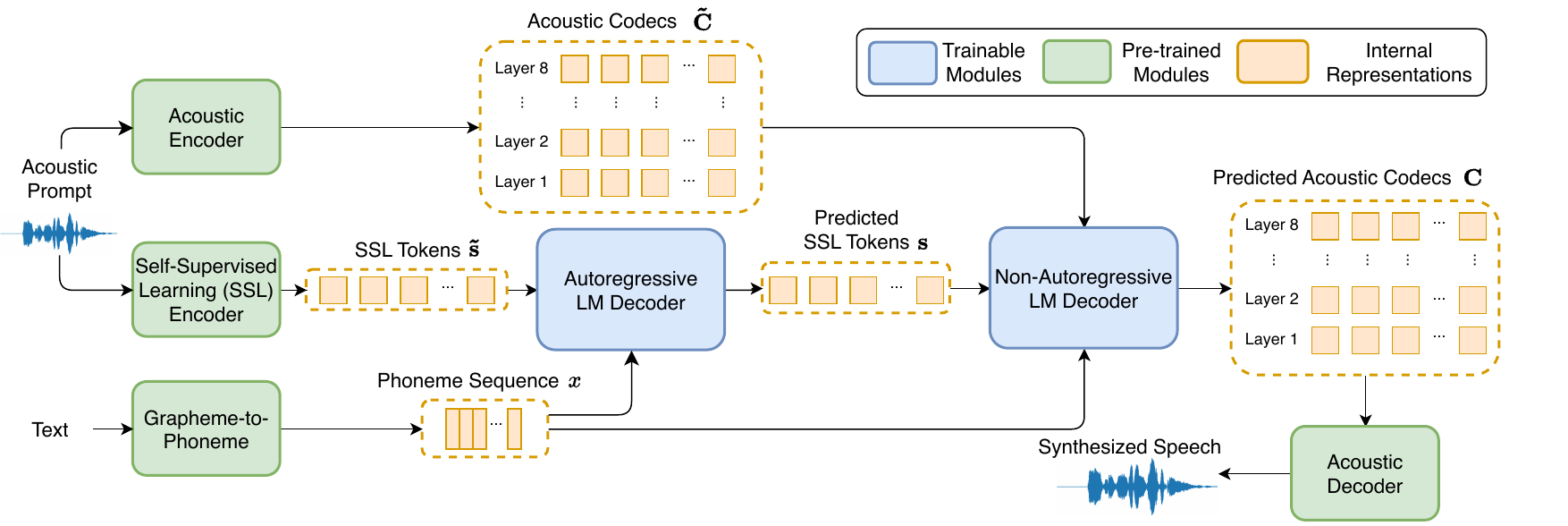}
    \caption{The overall diagram of proposed phonetic enhanced language model (LM) based text-to-speech framework. Given input text and an acoustic prompt, the autoregressive decoder predicts self-supervised learning (SSL) tokens that contain phonetic information, and the non-autoregressive decoder further predicts 8 layers of acoustic codecs that represent fine-grained acoustic details.}
    \label{fig:overall}
\end{figure*}

\section{Language-Model-Based TTS}

Text-to-Speech (TTS) models facilitate the translation between text and speech. TTS systems, such as Tacotron \cite{wang2017tacotron} and Transformer-based TTS \cite{li2019neural,okamoto2020transformer}, utilize sequence-to-sequence autoregressive models, significantly improving the naturalness of synthesized voices. However, the output sequences of TTS are continuous and often longer than text inputs, making it challenging to model abundant speech variations from discrete text inputs \cite{lei2023improving}. Incorrect predictions of text-acoustic alignment often result in robustness issues for synthesized voices \cite{okamoto2019tacotron}. Non-autoregressive TTS models, such as FastSpeech \cite{ren2019fastspeech} and Glow-TTS \cite{kim2020glow}, introduce a separate duration predictor, which improves TTS robustness but at the cost of reduced speech diversity and expressiveness \cite{tan2021survey}.

Inspired by the recent successes of prompting-based language models \cite{achiam2023gpt}, language model-based TTS \cite{wang2023neural, kharitonov2023speak, shen2023naturalspeech} has garnered attention. Specifically, neural audio codecs \cite{defossez2022high, zeghidour2021soundstream} facilitate the integration of speech units into a language model by quantizing speech waveforms into highly compressed audio codecs. TTS is then formulated as a task of conditional language modeling rather than as traditional signal regression.  VALL-E \cite{wang2023neural} represents one successful example of language model-based TTS, which demonstrates scalability and the capability for in-context learning. It employs a two-stage training process: modeling the first layer of audio codecs autoregressively and predicting the following layers in parallel.  VALL-E can be trained on large and diverse multi-speaker datasets and clone the voices of speakers from just a short acoustic prompt.  However, VALL-E inherits synthesis robustness issues, such as word mispronunciation, deletion, and repetition, from autoregressive language models \cite{wang2023neural}.

In this paper, we propose a phonetic enhanced language modeling method, aiming to facilitate a more robust language modeling for speech units.   The most related study, SPEAR-TTS \cite{kharitonov2023speak}, decouples TTS into two sequence-to-sequence tasks: from text to semantic tokens (`Reading') and from semantic tokens to speech waveforms (`Speaking'). The primary objective of SPEAR-TTS is to reduce the need for paired text and audio data, while our objective is to improve the robustness of language model-based TTS systems.

\section{Phonetic Enhanced Language Modeling}
In this section, we introduce our proposed phonetic enhanced language model (LM) based text-to-speech (TTS) system. Figure \ref{fig:overall} illustrates the overall diagram. The overall diagram follows the pipeline of ``Phoneme $\rightarrow$ Phonetic $\rightarrow$ Acoustic". Given phoneme inputs and an acoustic prompt as inputs, the autoregressive decoder first predicts SSL tokens that contain phonetic information. The non-autoregressive decoder then predicts 8 layers of acoustic codecs that represent fine-grained acoustic details. Finally, from these predicted acoustic codecs, the acoustic decoder synthesizes the speech waveform.

\subsection{Problem Formulation}

Following VALL-E \cite{wang2023neural}, we regard zero-shot TTS as a conditional language modeling task. Given phoneme sequences and an acoustic prompt, the TTS model is trained to generate discrete acoustic codecs. These codecs capture all the necessary acoustic details for synthesizing a speech waveform, which not only represent the linguistic content of the phoneme sequence but also reflect the speaker's characteristics as indicated by the acoustic prompt. Encodec \cite{defossez2022high} is often used to quantize acoustic prompt into 8 layers of discrete acoustic codecs. The first layer of codecs is predicted autoregressively, while the remaining layers are generated in parallel, based on the first layer of codecs. However, given the paramount importance of the first quantizer in the reconstruction process \cite{defossez2022high, zeghidour2021soundstream}, the autoregressive model is required to process an excessive amount of information. Consequently, this could lead to an accumulation of errors and result in robustness issues in synthesized speech. In response to this challenge, we propose a phonetic enhanced language model approach to help with the autoregressive training, which we will introduce next.

\subsection{Phonetic Enhanced Autoregressive Language Modeling}
Rather than directly mapping phoneme sequence to acoustic tokens, we propose an approach that maps phoneme sequence to phonetic tokens in an autoregressive manner, which we refer to as ``phonetic enhanced". Drawing inspiration from the success of self-supervised learning (SSL) representations in automatic speech recognition, we leverage SSL representations as phonetically rich targets for autoregressive training. The phonetic richness of SSL features arises from the models' training objective to capture the underlying structure of the audio signal \cite{mohamed2022self}. For instance, to accurately predict masked parts of speech, a SSL model requires a deep understanding of phonetics and the combinations of sounds in natural language \cite{wells2022phonetic}. Moreover, these models are trained on large and diverse datasets of speech, exposing them to a wide range of phonetic variations, including different accents and speaking styles. This exposure allows them to learn robust phonetic representations across various speakers and contexts \cite{de2022probing}.

We train the autoregressive language model to predict the SSL tokens from the phoneme sequences. 
Given a dataset $D = \{x_i, y_i\}$, where $y_i$ is the $i$-th audio sample and $x_i$ is the corresponding phoneme sequence, we train an autoregressive (AR) decoder-only language model to predict SSL tokens $\mathbf{s}$, which is conditioned  on the phoneme sequence $\mathbf{x}$ and the SSL tokens $\mathbf{\tilde{s}}$ extracted from the acoustic prompt:

\begin{equation}
    p(\mathbf{s} | \mathbf{x}, \mathbf{\tilde{s}}; \theta_{AR}) = \prod^{T}_{t=0} p(\mathbf{s}_t | \mathbf{s}_{<t},  \mathbf{\tilde{s}},\mathbf{x};\theta_{AR} )                                                                                                                                                          
\end{equation}
where $T$ is the sequence length of SSL tokens $\mathbf{\tilde{s}}$.

The AR model aims to predict SSL tokens from the input phoneme sequence. We refer this stage as ``Phoneme $\rightarrow$ Phonetic". Since SSL tokens not only encode contextual cues but also prosodic variations, the AR model learns to map phoneme sequence to a wide range of phonetic variations. 

\subsection{Non-Autoregressive Language Modeling}
The non-autoregressive (NAR) language model is trained to recover fine-grained acoustic details from phonetic variations, a stage we refer to as ``Phonetic $\rightarrow$ Acoustic". The NAR model is conditioned on the phoneme sequence $\mathbf{x}$, the predicted SSL tokens $\mathbf{{s}}$, and the acoustic codecs $\mathbf{\tilde{C}}$ extracted from the acoustic prompt:
\begin{equation}
        p(\mathbf{C} | \mathbf{x}, \mathbf{\tilde{C}}; \theta_{NAR}) = \prod^{8}_{j=1} p(\mathbf{c}_{:,j} | \mathbf{C}_{:,<j}, \mathbf{s},  \mathbf{\tilde{C}},\mathbf{x};\theta_{NAR} ) 
\end{equation}
The acoustic codecs $\mathbf{{C}}$ consist of 8 layers of discrete codec sequences, each containing fine-grained acoustic details that could be used to reconstruct a speech waveform. The amount of information decreases with each layer.

\subsection{Inference via Prompting}
During inference, an acoustic prompt is provided, which triggers the SSL encoder and acoustic encoder to generate SSL tokens and 8 layers of acoustic codecs, respectively. Subsequently, the AR decoder predicts SSL tokens from the input phoneme sequence and the SSL tokens. Then, the NAR decoder predicts acoustic codecs using the predicted SSL tokens, acoustic codecs from the acoustic prompt, and the phoneme sequence. Finally, the acoustic decoder synthesizes the speech waveform based on the predicted acoustic codecs.

\section{Experiments}
\subsection{Experimental Setup}
We perform all experiments on the LibriTTS dataset \cite{zen2019libritts}, which consists of around 600 hours of speech data with transcriptions at a 24kHz sampling rate from 2,456 speakers. We merge the `train-clean' and `train-other' subsets to train the TTS framework and evaluate its performance on the `test-clean' and `test-other' subsets, respectively. Note that the `test-other' subset contains more challenging recording conditions compared to the `test-clean'. The speakers from both `test-other' and `test-clean' subsets are unseen during training.

We evaluate the zero-shot performance of our TTS framework on the test sets of the LibriTTS dataset, where the framework is expected to synthesize speech mimicking the voice of a given speaker not seen during training. To set up the experiments, for each speaker in the test set, we randomly choose one speech sample from the same speaker to serve as the acoustic prompt, distinct from the speech to be synthesized, and then synthesize the speech based on the transcriptions. All the synthesized speech is within 3 to 10 seconds. All the demos are publicly available online\footnote{\textbf{Speech Demos}: https://demos46.github.io/IS24/}.

For comparison, we choose VALL-E \cite{wang2023neural} as the baseline framework. VALL-E utilizes a similar decoder-only language model (LM) and predicts the first layer of acoustic codecs autoregressively, followed by predicting the other layers non-autoregressively. Additionally, we compare different configurations using self-supervised learning (SSL) tokens. Specifically, we compare Hubert \cite{hsu2021hubert} and WavLM \cite{chen2022wavlm} features, followed by K-means clustering \cite{lloyd1982least} with the cluster sizes of 500 and 1024 for both Hubert and WavLM.

\subsection{Implementation Details}
We implemented VALL-E based on an unofficial release\footnote{https://github.com/lifeiteng/vall-e}. Our proposed framework shares a similar architecture with VALL-E. Both the autoregressive and non-autoregressive models consist of a transformer architecture consisting of 12 layers, 16 attention heads, an embedding dimension of 1024, a feed-forward layer dimension of 4096 and a dropout rate of 0.1. We use a 24kHz Encodec\footnote{https://huggingface.co/facebook/encodec\_24khz} as the acoustic encoder to compress the speech waveform into 8 layers of discrete acoustic codes, while the Encodec decoder is utilized to recover 8 layers of discrete acoustic codes to the speech waveform. To extract SSL tokens from the speech waveform, we use official versions of the Hubert\footnote{https://dl.fbaipublicfiles.com/hubert/hubert\_xtralarge\_ll60k\_finetune\_\\ls960.pt} and WavLM\footnote{https://huggingface.co/microsoft/wavlm-large} models. We obtain the features from the 6th Hubert layer and the 24th WavLM layer respectively, and cluster them using the K-means algorithm \cite{lloyd1982least}. We manually upsample the SSL tokens from 16kHz to 24kHz to align with acoustic codecs. We train the TTS framework on 4 NVIDIA TESLA V100 64GB GPUs for 1M steps. The other training configurations remain the same as those in the unofficial release\footnotemark[2] of VALL-E.

\subsection{Objective Evaluation}
We conduct objective evaluation to assess the robustness performance of our TTS framework. We use Whisper\footnote{https://huggingface.co/openai/whisper-large-v3} to generate the transcriptions from the synthesized speech and calculate word error rate (WER) and character error rate (CER) with the ground-truth transcriptions.

We have reported the objective results in Table \ref{table:objective}. Our proposed phonetic enhanced LM-based TTS framework consistently outperforms the baseline VALL-E on both `test-clean' and `test-other' on WER and CER metrics. Regarding different SSL configurations, we observe that the proposed framework with WavLM tokens clustered with a K-means cluster size of 1024 (`with WavLM (K = 1024)') achieves the best performance on both `test-clean' and `test-other', while WavLM with a cluster size of 500 (`with WavLM (K = 500)') performs slightly worse than the others. Compared to WavLM, the cluster size of Hubert features does not have a significant impact on the final results. However, their performance declines on the `test-other', indicating that Hubert tokens are less noise-robust than WavLM tokens. 

\begin{table}[t]
\centering
\caption{Objective evaluation results to compare the robustness of the baseline VALL-E and our proposed TTS systems with  different configurations using SSL tokens. `WER' and `CER' denote as `Word Error Rate' and `Character Error Rate' respectively, where a smaller value indicates a better system robustness. `K' represents the K-means cluster size of the SSL tokens. }
\scalebox{0.9}{
\begin{tabular}{cc|cc|cc}
\hline
\multicolumn{2}{c|}{Test Dataset}                                                                                                                                                               & \multicolumn{2}{c|}{\begin{tabular}[c]{@{}c@{}}LibriTTS\\ (test-clean)\end{tabular}} & \multicolumn{2}{c}{\begin{tabular}[c]{@{}c@{}}LibriTTS\\ (test-other)\end{tabular}} \\ \hline
\multicolumn{2}{c|}{Objective Metrics (\%)}                                                                                                                                                      & \multicolumn{1}{c}{WER$\downarrow$}                            & CER$\downarrow$                         & \multicolumn{1}{c}{WER$\downarrow$}                            & CER$\downarrow$                        \\ \hline
\multicolumn{2}{c|}{\begin{tabular}[c]{@{}c@{}}Baseline\\ (VALL-E)\end{tabular}}                                                                                                                & \multicolumn{1}{c}{21.05}                          & 18.19                          & \multicolumn{1}{c}{25.56}                               &     23.51                       \\ \hline
\multicolumn{1}{c|}{\multirow{5}{*}{\begin{tabular}[c]{@{}c@{}}\\Proposed\\ Phonetic \\ Enhanced\\ LM-based\\ TTS\end{tabular}}} & \begin{tabular}[c]{@{}c@{}}w/ Hubert\\ (K = 500)\end{tabular}  & \multicolumn{1}{c}{8.03}                           & 7.18                           & \multicolumn{1}{c}{13.39}                          & 11.58                          \\ \cline{2-6} 
\multicolumn{1}{c|}{}                                                                                                          & \begin{tabular}[c]{@{}c@{}}w/ Hubert\\ (K = 1024)\end{tabular} & \multicolumn{1}{c}{7.66}                           & 5.84                           & \multicolumn{1}{c}{12.14}                          & 10.05                          \\ \cline{2-6} 
\multicolumn{1}{c|}{}                                                                                                          & \begin{tabular}[c]{@{}c@{}}w/ WavLM\\ (K = 500)\end{tabular}   & \multicolumn{1}{c}{12.72}                           & 11.96                          & \multicolumn{1}{c}{14.34}                          & 12.73                          \\ \cline{2-6} 
\multicolumn{1}{c|}{}                                                                                                          & \begin{tabular}[c]{@{}c@{}}w/ WavLM\\ (K = 1024)\end{tabular}  & \multicolumn{1}{c}{6.12}                           & 5.44                           & \multicolumn{1}{c}{7.92}                           & 6.11                           \\ \hline
\end{tabular}}
\label{table:objective}
\end{table}

\begin{table}[t]
\centering
\caption{Subjective evaluation results with 95\% confidence interval to assess the naturalness (`MOS') and the speaker similarity (`S-MOS') of the baseline VALL-E and our proposed TTS systems with different configurations using SSL tokens.}
\scalebox{0.76}{\begin{threeparttable}[t]
\begin{tabular}{cc|cc|cc}
\hline
\multicolumn{2}{c|}{Test Dataset}                                                                                                                                                               & \multicolumn{2}{c|}{\begin{tabular}[c]{@{}c@{}}LibriTTS\\ (test-clean)\end{tabular}} & \multicolumn{2}{c}{\begin{tabular}[c]{@{}c@{}}LibriTTS\\ (test-other)\end{tabular}} \\ \hline
\multicolumn{2}{c|}{\begin{tabular}[c]{@{}c@{}}Subjective\\ Metrics\end{tabular}}                                                                                                                                                         & \multicolumn{1}{c|}{MOS}                           & S-MOS                           & \multicolumn{1}{c|}{MOS}                           & S-MOS                           \\ \hline
\multicolumn{2}{c|}{\begin{tabular}[c]{@{}c@{}}Reference\\ Speech\end{tabular}}                                                                                                                                                                  & \multicolumn{1}{c|}{4.81$\pm$0.07 }                              &                $-^*$                 & \multicolumn{1}{c|}{4.39$\pm$0.11 }                              &                $-^*$                \\ \hline
\multicolumn{2}{c|}{\begin{tabular}[c]{@{}c@{}}Baseline\\ (VALL-E)\end{tabular}}                                                                                                                & \multicolumn{1}{c|}{2.51$\pm$0.19}                              &           3.12$\pm$0.17                      & \multicolumn{1}{c|}{2.49$\pm$0.16}                              &           3.09$\pm$0.13                     \\ \hline
\multicolumn{1}{c|}{\multirow{4}{*}{\begin{tabular}[c]{@{}c@{}}\\Proposed\\ Phonetic \\ Enhanced\\ LM-based\\ TTS\end{tabular}}} & \begin{tabular}[c]{@{}c@{}}w/ Hubert\\ (K = 500)\end{tabular}  & \multicolumn{1}{c|}{3.58$\pm$0.13}                              &       3.89$\pm$0.13                          & \multicolumn{1}{c|}{3.43$\pm$0.13}                              &         3.35$\pm$0.14                      \\ \cline{2-6} 
\multicolumn{1}{c|}{}                                                                                                          & \begin{tabular}[c]{@{}c@{}}w/ Hubert\\ (K = 1024)\end{tabular} & \multicolumn{1}{c|}{3.29$\pm$0.14}                              &      3.59$\pm$0.14                           & \multicolumn{1}{c|}{3.19$\pm$0.10}                              &      3.33$\pm$0.15                           \\ \cline{2-6} 
\multicolumn{1}{c|}{}                                                                                                          & \begin{tabular}[c]{@{}c@{}}w/ WavLM\\ (K = 500)\end{tabular}   & \multicolumn{1}{c|}{3.38$\pm$0.13}                              &     3.61$\pm$0.16                           & \multicolumn{1}{c|}{3.33$\pm$0.14}                              &         3.53$\pm$0.13                        \\ \cline{2-6} 
\multicolumn{1}{c|}{}                                                                                                          & \begin{tabular}[c]{@{}c@{}}w/ WavLM\\ (K = 1024)\end{tabular}  & \multicolumn{1}{c|}{3.54$\pm$0.13}                              &             3.94$\pm$0.13                         & \multicolumn{1}{c|}{3.53$\pm$0.10}                              &      3.55$\pm$0.15                     \\ \hline
\end{tabular}
  \begin{tablenotes}
     \item[*] S-MOS results for `Reference Speech' are not reported because it serves as the comparison target in speaker similarity evaluation.
   \end{tablenotes}
   \end{threeparttable}}
\label{tab:mos}
\vspace{-4mm}
\end{table}

\subsection{Subjective Evaluation}
We further conduct two listening tests to assess the performance of our TTS framework. In each test, 10 native English speakers participated, with each participant listening to a total of 200 synthesized speech samples (200 samples = 10 samples $\times$ 5 systems $\times$ 2 test sets $\times$ 2 tests). The first listening test (referred as `MOS') evaluates the naturalness of the synthesized speech samples. Participants are asked to score each sample on a 5-point scale (5: Excellent, 4: Good, 3: Satisfactory, 2: Unsatisfactory, 1: Bad). The second listening test (referred as `S-MOS') evaluates the speaker similarity with the reference speech. Participants are asked to score each sample on a 5-point scale (5: Very Similar, 4: Similar, 3: Neutral, 2: Dissimilar, 1: Very Dissimilar). The results are reported in Table \ref{tab:mos}.

We first observe that all of our proposed frameworks significantly outperform the Baseline (VALL-E) in  naturalness and speaker similarity on both test sets (`test-clean' and `test-other'). This demonstrates that our phonetic enhanced language modeling approach improves the robustness of the TTS system and contributes to better zero-shot performance for synthesized speech.

We further investigate the impacts of different configurations of SSL tokens. From Table \ref{tab:mos}, we observe that the proposed framework with Hubert features clustered with a size of 500 (`w/ Hubert (K=500)') outperforms that with a size of 1024 (`w/ Hubert (K=1024)') on the test set of LibriTTS (test-clean). Conversely, the proposed framework with WavLM features clustered with a size of 1024 (`w/ WavLM (K=1024)') outperforms that with a size of 500 (`w/ WavLM (K=500)') on the same test set. Additionally, we note that the proposed (`w/ WavLM (K=1024)') achieves slightly better speaker similarity than the proposed (`w/ Hubert (K=500)'), while the naturalness slightly declines. 

We observe that all frameworks suffers from a performance decline on the test set of LibriTTS (test-other), which contains more challenging recording conditions. 
Notebly, the performance of the proposed frameworks with WavLM tokens presents a more noise-robust performance by achieving less performance decline than those with Hubert tokens.
\subsection{Discussions}
The major difference between our proposed TTS framework and the baseline VALL-E is the utilization of self-supervised representations as phonetic targets for autoregressive language modeling. Both objective and subjective evaluations reveal two key findings:
\begin{itemize}
    \item The proposed TTS framework utilizing WavLM tokens demonstrates greater noise robustness compared to those using HuBERT tokens;
    \item A larger cluster size in K-means clustering of WavLM tokens contributes to improved performance.
\end{itemize}
As a result, configurations using WavLM features with a larger cluster size outperform other setups. This advantage is largely credited to WavLM's pre-training strategy, which includes denoising and handling distorted audio signals, thereby enhancing its ability to capture phonetic details under challenging conditions. WavLM leverages more detailed phonetic information provided by a larger cluster and better model the nuances in complex audio environments. 
However, increasing the cluster size in Hubert could potentially add complexity to non-autoregressive acoustic modeling and might lead to diminished performance.

Our future improvement focuses on the acoustic encoder, for which we use Encodec \cite{defossez2022high} in our experiments. We found that the performance of our proposed TTS system is primarily limited by the effectiveness of Encodec, particularly in terms of its reconstruction distortion, noise robustness, codebook size, and compression bitrate.

\section{Conclusion}
This paper proposes a phonetic enhanced language modeling method for text-to-speech synthesis. By leveraging self-supervised representations as the training target for the autoregressive language model and predicting fine-grained acoustic details non-autoregressively, the text-to-speech framework learns to map phoneme sequences to phonetic variations and then to the acoustic information. Our experiments demonstrate that the proposed method could effectively reduce error propagation in speech unit modeling and improve the robustness and naturalness of the synthesized speech. Our experiments also compare different configurations of self-supervised representations and analyze their impacts on the synthesized results.

\newpage
\bibliographystyle{IEEEtran}
\bibliography{mybib}

\end{document}